\documentclass[11pt]{article}
\usepackage[utf8]{inputenc}
\usepackage{amsmath, graphicx, subcaption}
\begin{document}
\title{Universal Self-Similar Attractor in the Bending-Driven Leveling of Thin Viscous Films}
\author{Christian Pedersen$^1$ \and Thomas Salez$^{2,3,\star}$ \and Andreas Carlson$^1$}
\date{
\scriptsize $^1$Mechanics Division, Department of Mathematics, University of Oslo, 0316 Oslo, Norway.\\
$^2$Univ. Bordeaux, CNRS, LOMA, UMR 5798, F-33405 Talence, France.\\
$^3$Global Station for Soft Matter, Global Institution for Collaborative Research and Education, Hokkaido University, Sapporo, Hokkaido 060-0808, Japan.\\
$^\star$thomas.salez@u-bordeaux.fr\\
\normalsize\today
}
\maketitle

\begin{abstract}
We study theoretically and numerically the bending-driven leveling of thin viscous films within the lubrication approximation. We derive the Green's function of the linearized thin-film equation and further show that it represents a universal self-similar attractor at long times. As such, the rescaled perturbation of the film profile converges in time towards the rescaled Green's function, for any summable initial perturbation profile. In addition, for stepped axisymmetric initial conditions, we demonstrate the existence of another, short-term and one-dimensional-like self-similar regime. Besides, we characterize the convergence time towards the long-term universal attractor in terms of the relevant physical and geometrical parameters, and provide the local hydrodynamic fields and global elastic energy in the universal regime as functions of time. Finally, we extend our analysis to the non-linear thin-film equation through numerical simulations.
\end{abstract}

\section*{Introduction}

Confined elastohydrodynamic flows have been of great interest to the scientific community~\cite{kumar201870,juel2018instabilities} due to their ubiquitous appearance in Nature~\cite{parau2002nonlinear,michaut2011dynamics, puaruau2011three}, physiology~\cite{grotberg2004biofluid}, and
industry~\cite{christov2018flow}.
A particular case of such flows corresponds to an elastic plate supported by a thin viscous fluid film. When subjected to a perturbation, this setting can exhibit a wide range of dynamical behaviours depending on the system size, its geometry, the boundary conditions, and the perturbation itself.
Examples encompass viscous fingering instabilities~\cite{pihler2012suppression, pihler2013modelling, al2013two,  pihler2014interaction, peng2019viscous}, wrinkling and buckling~\cite{Vandeparre2010,kodio2017lubricated,huang2002wrinkling}, elastocapillary rise~\cite{duprat2011dynamics}, dewetting-like rim formation due to attractive van der Waals forces~\cite{carlson2016similarity}, biologically-induced membrane adhesion~\cite{leong2010adhesive, carlson2015protein}, flow induced by applied pressure fields~\cite{elbaz2016axial}, including wake formation~\cite{arutkin2017elastohydrodynamic} and peristaltic flow in cylindrical geometries~\cite{takagi2011peristaltic, romano2020peristaltic}.

At the core of the above phenomena one invariably finds the coupling between the elastic forces in the plate, \textit{i.e.} bending or in-plane tension, and the viscous forces in the supporting fluid. A canonical situation to study such a coupling is the spreading of an excess volume of fluid under an elastic plate. This is the elastohydrodynamic analog to the famous capillary-driven spreading of a liquid droplet revealed by Tanner~\cite{tanner1979spreading}. When theoretically describing the spreading of a liquid droplet on a solid substrate, one needs to consider with care the contact-line region where the film thickness becomes vanishingly small and the viscous dissipation should diverge~\cite{huh1971hydrodynamic} -- despite our common observation of moving droplets on windows. To overcome the latter paradox, several regularisation mechanisms have been proposed, such as slippage~\cite{huh1971hydrodynamic}, the formation of a vapour tip at the advancing front with the fluid lagging behind~\cite{hewitt2015elastic, ball2018static, berhanu2019uplift}, diffusion ~\cite{carlson2009pof} or the existence of a nanometric precursor film over which the droplet spreads~\cite{hosoi2004peeling, lister2013viscous}.

Therefore, the precursor film situation is emblematic, and deserves to be revisited in an elastohydrodynamic context, \textit{e.g.} with an elastic plate replacing the free interface of the droplet. For small plate deflections, or spatially-unconstrained plates~\cite{pedersen2019asymptotic}, the viscous flow is solely driven by the bending forces. In the case of a small thickness of the precursor film relative to the droplet height, the spreading law can be obtained by asymptotic matching between a quasi-static solution and a traveling-wave solution at the advancing front~\cite{lister2013viscous}. However, if the droplet volume is constant, the spreading process will inevitably lead to a situation in which the height of the film perturbation becomes less than that of the precursor film. This asymptotic regime, before the plate settles into its flat equilibrium shape, corresponds to a linear thin-film equation and it is thus tractable using traditional mathematical methods~\cite{fourier1878analytical, green1889essay}. Nevertheless, there is still a physical interest about the characteristic traits describing the solution in this regime.

The dynamics of such systems is often described within the framework of lubrication theory~\cite{batchelor2000introduction,stillwagon1988,oron1997long,Craster2009,Blossey2012}, which for small plate deformations leads to the bending-driven thin-film equation.
Since thin-film equations are high-order non-linear diffusive-like equations, they have attracted substantial attention from the applied mathematical community~\cite{witelski1998self, Bertozzi1998,  carrillo2002long, biler2002intermediate, witelski2003adi, galaktionov2004intermediate,  Munch2005, bowen2006linear, bartier2011improved, chapman2013exponential, huang2013group, giacomelli2016rigorous, witelski2020nonlinear, christov2020long, segatti2020fractional}.
In particular, the general solution in terms of the heat kernel to the $n$-th order linear equation has been presented in ~\cite{barvinsky2019heat}, providing a detailed discussion on its increasingly oscillatory behaviour with increasing $n$, and its asymptotic nature, which has been applied in \textit{e.g.} quantum electrodynamics~\cite{gusynin2020landau}. Similar studies on general solutions to the $n$-th order equation have also been performed using the Green's function associated with high-order differential operators ~\cite{avramidi1998green}.
For the specific case of bending-driven elastohydrodynamic flows, the literature is scarce compared to its capillary analogue. Yet, there are notable advances that have been made. Several solutions valid for specific initial boundary value problems have been found using asymptotic and numerical methods ~\cite{flitton2004moving}, highlighting the solution sensitivity to the fluid boundary conditions. Also, the existence of weak solutions has been shown using appropriate boundary conditions~\cite{bernis1990higher, li2012propagation, korzec2012global} for which convergence has been demonstrated using a finite-element approximation ~\cite{barrett2004finite}.
However, as there exists no general analytical solution to such non-linear equations, intermediate asymptotics~\cite{barenblatt1996scaling} and self-similar solutions~\cite{evans2007unstable, galaktionov2008countable, sekimoto2019symmetry} have proven to be essential tools for revealing a degree of generality. As an illustration, a recent study on the intermediate asymptotics of the capillary-driven thin-film equation has shown that any vanishingly-small and summable initial perturbation profile at the free surface of the film must converge in time towards a universal self-similar attractor, given by the Green's function of the linearized problem~\cite{benzaquen2013intermediate}. This convergence scenario was then explored experimentally using non-flat molten polymer nanofilms \cite{Baumchen2013,benzaquen2014approach,Backholm2014} and droplet bouncing on oil layers~\cite{lakshman2021deformation}. It was also generalized to higher-order symmetries of the initial profile with practical implications regarding stability and dynamics for laser lithography~\cite{Benzaquen2015}.

In the present article, we investigate theoretically and numerically the intermediate asymptotics of the bending-driven thin-film equation for a Newtonian fluid. In particular, for any summable initial perturbation with respect to the flat equilibrium state, we show that the solution converges towards a universal self-similar attractor provided by the Green's function of the linearized problem.  In addition, for stepped axisymmetric initial conditions, we demonstrate the existence of another, early-time and one-dimensional-like self-similar regime. Besides the obvious material parameters of the elastic plate (bending stiffness) and the liquid (viscosity), we show that the convergence time towards the long-term universal attractor is essentially set by the sixth power of the typical width of the initial perturbation. We also derive the hydrodynamic fields and elastic energy in the universal regime as functions of time. Finally, using numerical methods, we extend these results to the case of finite perturbations.

\section*{Physical model}
We consider a free unconstrained elastic plate, with zero spontaneous curvature, resting on a thin liquid film, \textit{i.e.} the lateral extent of any variation of the film profile is much larger than the film thickness itself, as required for the lubrication approximation to be valid~\cite{reynolds1895iv}. The liquid is described as an incompressible and Newtonian viscous fluid, that is supported by a solid substrate at the vertical coordinate $\bar{z}=0$. At time $\bar{t}$, and horizontal coordinates $(\bar{x}, \bar{y})$, the liquid-plate interface is located at $\bar{z}=\bar{h}(\bar{x}, \bar{y}, \bar{t})$, where $\bar{h}(\bar{x}, \bar{y}, \bar{t})$ thus denotes the interface profile. In the lubrication approximation, the excess pressure field $\bar{p}(\bar{x}, \bar{y}, \bar{t})$ in the liquid (with respect to the atmospheric pressure) is independent of $\bar{z}$. By continuity of the normal stress at the liquid-plate interface, which is in this case set by the elastic stress due to the bending of the plate~\cite{landau1959course}, we get:
\begin{equation}
\bar{p}(\bar{x}, \bar{y}, \bar{t}) = B\bar{\nabla}^4\bar{h}(\bar{x}, \bar{y}, \bar{t})\ ,
\label{eq:pressure}
\end{equation}
where $B$ is the plate's bending stiffness and $\bar{\nabla}$ the nabla operator in the $\bar{x}\bar{y}$-plane. By solving the Stokes equations within the lubrication approximation~\cite{batchelor2000introduction}, under no-slip boundary conditions at the two solid-liquid interfaces, we find the horizontal velocity field in the liquid:
\begin{equation}
\bar{u}(\bar{x}, \bar{y}, \bar{z}, \bar{t}) = \frac{1}{2\mu}\left(\bar{z}^2-\bar{h}\bar{z}\right)\bar{\nabla} \bar{p}(\bar{x}, \bar{y}, \bar{t}) \ .
\label{eq:velocity}
\end{equation}
By integrating the continuity equation across the film thickness, we obtain the bending-driven thin-film equation~\cite{lister2013viscous,pedersen2019asymptotic}:
\begin{equation}
\partial_{\bar{t}} \bar{h}(\bar{x}, \bar{y}, \bar{t}) = \frac{B}{12\mu}\bar{\nabla}\cdot\left[\bar{h}^3(\bar{x}, \bar{y}, \bar{t})\bar{\nabla}^5 \bar{h}(\bar{x}, \bar{y}, \bar{t}) \right]\ .
\label{eq:ehd_dimensional}
\end{equation}
We nondimensionalize Eq.~\eqref{eq:ehd_dimensional} with $\bar{h} = hh_0$, $\bar{x}=xh_0$, $\bar{y}=yh_0$ and $\bar{t}=12t\mu h_0^3 / B$, where $h_0$ is the liquid film thickness in the flat unperturbed state. By doing so, we obtain the dimensionless version of Eq.~\eqref{eq:ehd_dimensional}:
\begin{equation}
\partial_t h(x,y,t) = \nabla\cdot\left[h^3(x,y,t)\nabla^5 h(x,y,t)\right]\ ,
\label{eq:ehd_nondimensional}
\end{equation}
where $\nabla$ is the dimensionless nabla operator in the $xy$-plane.

\section*{Linearized problem}
We consider small perturbations of the film height, \textit{i.e.} $h(x, y, t) = 1 + \epsilon(x, y, t)$ with $\epsilon \ll 1$, so that Eq.~\eqref{eq:ehd_nondimensional} can be linearized into:
\begin{equation}
\partial_t \epsilon(x,y,t) = \nabla^6 \epsilon(x,y,t)\; .
\label{eq:ehd_linear}
\end{equation}

\subsection*{Green's function and symmetries}
The Green's function $G(x,y,t)$ is defined as the solution of the following partial differential equation:
\begin{equation}
\mathcal{L}G(x,y,t) = \delta(x,y,t)\ ,
\label{eq:Greens}
\end{equation}
where $\mathcal{L} = \partial_t - (\partial_x^2 + \partial_y^2)^3$ is the linear differential operator of Eq.~\eqref{eq:ehd_linear}, and $\delta(x,y,t)$ is the Dirac delta function in two-dimensional space and time. The solution $\epsilon(x,y,t)$, at any position ($x$, $y$) and time $t$, is then obtained from a convolution between the Green's function and the initial profile
$\epsilon_0(x,y)=\epsilon(x,y,0)$, as:
\begin{equation}
\epsilon(x,y,t) = \int \textrm{d}x'\,\textrm{d}y'\, G(x-x', y-y', t)\epsilon_0(x',y')\ .
\label{eq:convolution}
\end{equation}

We invoke the Fourier transform:
\begin{equation}
\hat{G}(k_x,k_y,\omega) = \int \textrm{d}x\,\textrm{d}y\, \textrm{d}t\, G(x,y,t)e^{-i(k_xx + k_yy + \omega t)}\ ,
\end{equation}
with $k_x$ and $k_y$ the angular spatial frequencies in the $x$ and $y$ directions, respectively, and $\omega$ the angular temporal frequency. By taking the Fourier transform of Eq.~\eqref{eq:Greens}, we find:
\begin{equation}
\hat{G}(k_x, k_y, \omega) = \frac{1}{i\omega + \left(k_x^2 + k_y^2\right)^3}\ .
\end{equation}
Expressing the inverse Fourier transform, and invoking the residue theorem for the integral over the angular temporal frequency, we obtain the Green's function in a general integral form:
\begin{equation}
G(x, y, t) = \frac{\mathcal{H}(t)}{(2\pi)^2}\int \textrm{d}k_x\,\textrm{d}k_y\, e^{-t\left(k_x^2+k_y^2\right)^3}e^{i(k_xx + k_yy)}\ ,
\label{eq:greens_integral}
\end{equation}
where $\mathcal{H}(t)$ is the Heaviside step function.

Let us now perform a change of variables towards polar coordinates, through $x=r\cos(\theta)$, $y=r\sin(\theta)$, $k_x=\rho\cos(\psi)$ and $k_y=\rho\sin(\psi)$, where $\theta$ and $\psi$ are the azimuthal and polar angles, respectively. When inserted into Eq.~\eqref{eq:greens_integral}, this change of variables leads to:
\begin{equation}
\begin{split}
G(r, t) &= \frac{\mathcal{H}(t)}{(2\pi)^2}\int \textrm{d}\rho\, \rho e^{-\rho^6t} \int \textrm{d}\psi\, e^{i\rho r\cos(\psi - \theta)} \ , \\
        &= \frac{\mathcal{H}(t)}{2\pi}\int \textrm{d}\rho\, \rho e^{-\rho^6t} J_0(\rho r)\ ,
\end{split}
\end{equation}
with $J_0$ the zeroth-order Bessel function. As a consequence, the Green's function is axisymmetric. Furthermore, the last integral has an exact expression, leading to:
\begin{equation}
\begin{split}
G(r,t) = &\frac{\mathcal{H}(t)\Gamma\left(\frac{1}{3}\right)}{12\pi t^{1/3}} {}\,_0F_4\left[ \left\{\frac{1}{3}, \frac{2}{3}, \frac{2}{3}, 1 \right\}, -\left( \frac{1}{6}\frac{r}{t^{1/6}} \right)^6  \right]\\ &+ \frac{\mathcal{H}(t)r^2\Gamma\left(-\frac{1}{3}\right)}{144\pi t^{2/3}}{}\,_0F_4\left[ \left\{ \frac{2}{3}, 1, \frac{4}{3}, \frac{4}{3} \right\}, -\left( \frac{1}{6}\frac{r}{t^{1/6}}  \right)^6  \right]\\ &+ \frac{\mathcal{H}(t)r^4}{768\pi t} {}\,_0F_4\left[ \left\{ \frac{4}{3}, \frac{4}{3}, \frac{5}{3}, \frac{5}{3} \right\}, -\left( \frac{1}{6}\frac{r}{t^{1/6}} \right)^6  \right]\ ,
\label{eq:Greens_function}
\end{split}
\end{equation}
where $\Gamma$ is the gamma function and $_0F_4$ is the (0,4)-hypergeometric function. We note that the Green's function in Eq.\eqref{eq:Greens_function} is similar to the one obtained in~\cite{tulchinsky2016transient}.

Finally, introducing the similarity variable $\xi = r t^{-1/6}$, Eq.~\eqref{eq:Greens_function} can be recast into:
\begin{equation}
G(\xi, t) = \frac{\mathcal{H}(t)}{t^{1/3}}f(\xi)\ ,
\label{ss}
\end{equation}
where:
\begin{equation}
\begin{split}
f(\xi) = &\frac{\Gamma\left(\frac{1}{3}\right)}{12\pi} {}_0F_4\left( \left\{\frac{1}{3}, \frac{2}{3}, \frac{2}{3}, 1 \right\}, -\left( \frac{\xi}{6} \right)^6  \right)\\ &+ \frac{\xi^2\Gamma\left(-\frac{1}{3}\right)}{144\pi}{}_0F_4\left( \left\{ \frac{2}{3}, 1, \frac{4}{3}, \frac{4}{3} \right\}, -\left( \frac{\xi}{6}  \right)^6  \right)\\ &+ \frac{\xi^4}{768\pi} {}_0F_4\left( \left\{ \frac{4}{3}, \frac{4}{3}, \frac{5}{3}, \frac{5}{3} \right\}, -\left( \frac{\xi}{6} \right)^6  \right)\ .
\label{eq:attractor}
\end{split}
\end{equation}
As a consequence, one has:
\begin{equation}
\frac{G(\xi, t)}{G(0, t)} = \frac{f(\xi)}{f(0)}\ ,
\label{eq:normalized_attractor}
\end{equation}
which means that, when properly normalized, the Green's function is essentially a function of the self-similar variable $\xi$ only.

\subsection*{General solution and long-term behaviour}
In general, the double integral of Eq.~\eqref{eq:greens_integral} can be evaluated numerically, as well as the solution of Eq.~\eqref{eq:ehd_linear} for any initial profile, using Eq.~\eqref{eq:convolution}. Moreover, in the particular case where the initial profile is axisymmetric, \textit{i.e.} $\epsilon_0(x,y)=\epsilon_0(r)$, the spatial convolution defined in Eq.~\eqref{eq:convolution} reads in polar coordinates:
\begin{equation}
\epsilon(r,t) = \int \textrm{d}r'\, r'\epsilon_0(r')\int \textrm{d}\theta\, G\left(\sqrt{r^2+{r'}^2 - 2rr'\cos(\theta)},t\right)\ .
\label{eq:convolution_axi}
\end{equation}
Therefore, the solution is axisymmetric at any time, as expected.

\begin{figure}[th!]
\begin{subfigure}[b]{0.52\linewidth}
\includegraphics[width=1\linewidth]{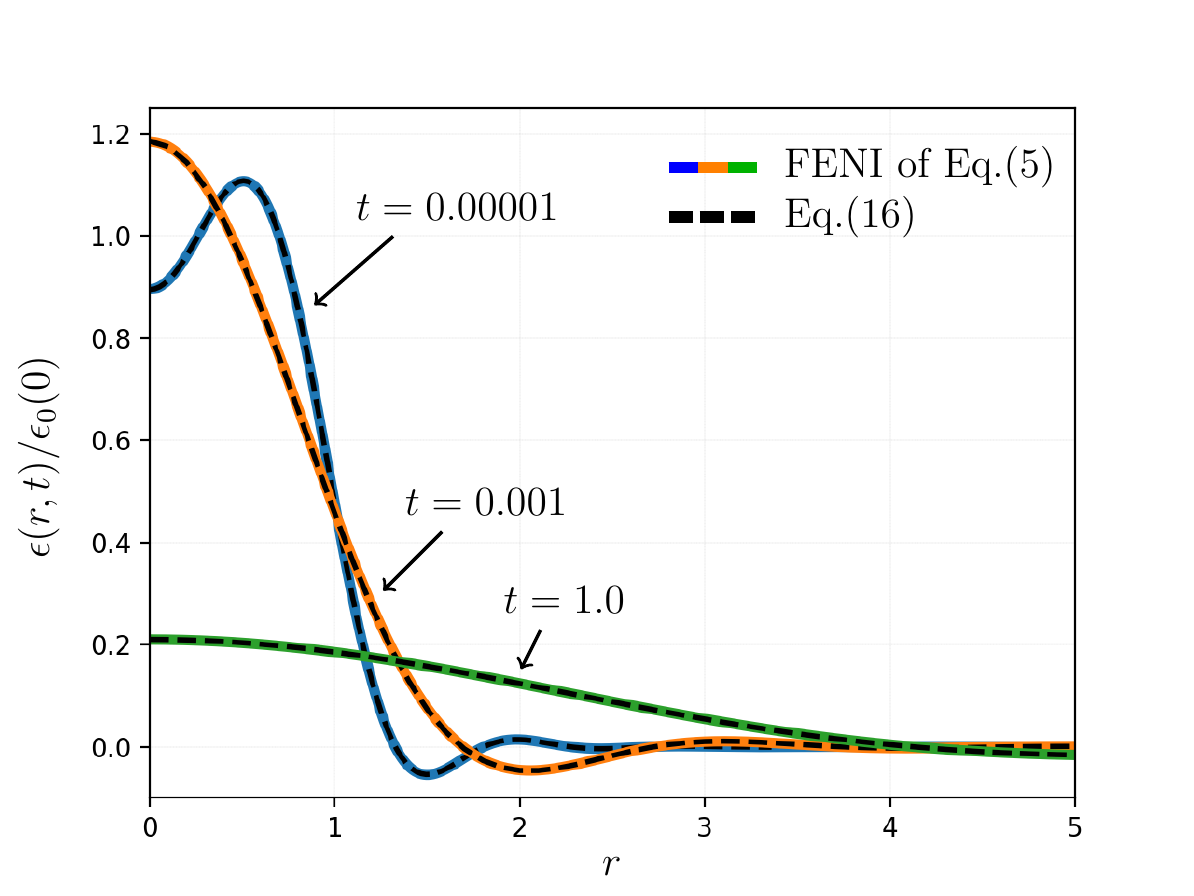}
\caption{}
\end{subfigure}
\begin{subfigure}[b]{0.52\linewidth}
\includegraphics[width=1\linewidth]{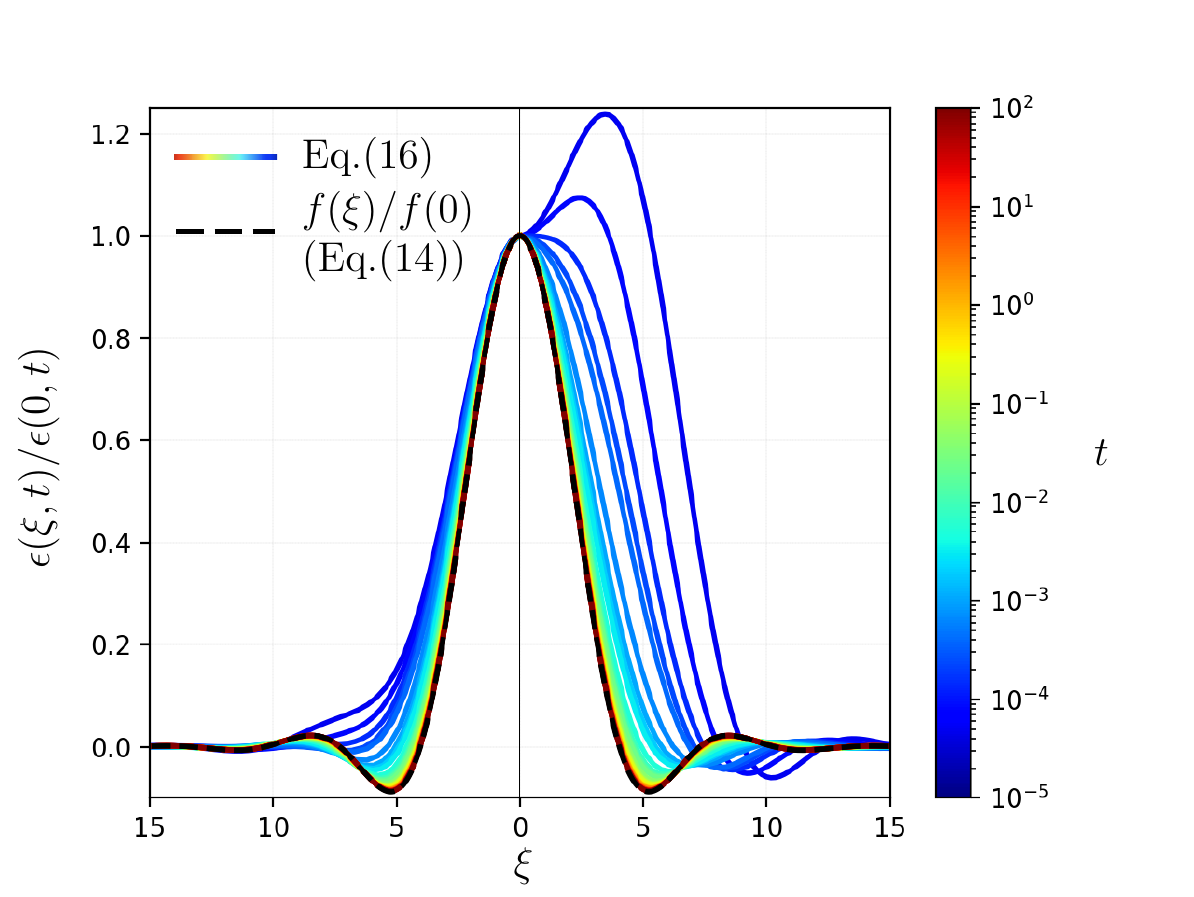}
\caption{}
\end{subfigure}
\caption{(a) Normalized solutions $\epsilon(r,t)/\epsilon_0(0)$ as a function of the radial coordinate $r$, at three different times $t$ as indicated, for a stepped axisymmetric initial profile $\epsilon_0(r) = \mathcal{H}(1-r)$. These solutions were obtained from: i) finite-element numerical integration (FENI)~\cite{pedersen2019asymptotic} of Eq.~\eqref{eq:ehd_linear} (dashed lines); ii) numerical evaluation of the convolution in Eq.~\eqref{eq:convolution_axi} (solid lines). (b) Rescaled solutions $\epsilon(r,t)/\epsilon(0,t)$ as a function of the similarity variable $\xi=rt^{-1/6}$ (solid lines), for various times $t$ (color bar), as numerically computed from Eq.~\eqref{eq:convolution_axi}, for two different axisymmetric initial profiles: i) an homogeneous polynomial $\epsilon_0(r)=\left(1-r^2\right)^2\mathcal{H}(1-r)$ (left) ; ii) a stepped axisymmetric function $\epsilon_0(r) = \mathcal{H}(1-r)$ (right). For comparison, $f(\xi)/f(0)$ (see Eq.~\eqref{eq:attractor}) is shown (dashed line).}
\label{fig:exact_solution}
\end{figure}

In Fig.~\ref{fig:exact_solution}a, we compare the solution of a finite-element numerical integration (FENI)~\cite{pedersen2019asymptotic} of Eq.~\eqref{eq:ehd_linear} and the numerical evaluation of the convolution in Eq.~\eqref{eq:convolution_axi}, for a stepped axisymmetric initial profile $\epsilon_0(r) = \mathcal{H}(1-r)$, at three different times $t$. We observe an excellent agreement, which confirms the validity of both the Green's function and the convolution.

As the magnitude of the axisymmetric solution $\epsilon(r,t)$ above decays with time (see Fig.~\ref{fig:exact_solution}a), to study the long-term behaviour
of the solution we rescale $\epsilon(r,t)$ by its amplitude $\epsilon(0,t)$ at $r=0$. Furthermore, guided by the self-similarity of the Green's function (see Eq.~\eqref{ss}), we introduce the similarity variable $\xi=rt^{-1/6}$ and study $\epsilon(\xi,t)/\epsilon(0,t)$. The latter rescaled solution is numerically computed from Eq.~\eqref{eq:convolution_axi} for two different axisymmetric initial profiles, and plotted in Fig.~\ref{fig:exact_solution}b as a function of $\xi$ for different points in time. In the left panel, we have used an homogeneous polynomial, $\epsilon_0(r)=(1-r^2)^2\,\mathcal{H}(1-r)$, as the initial profile; and in the right panel, we have used the stepped axisymmetric initial profile $\epsilon_0(r) = \mathcal{H}(1-r)$ previously employed in Fig.~\ref{fig:exact_solution}a. In both cases, the rescaled solutions appear to be independent of time $t$ at long times, which suggests their late-time self-similarity. Moreover, they both seem to converge towards $f(\xi)/f(0)$, suggesting the existence of a universal self-similar attractor.

\subsection*{Early-time similarity solution}

\begin{figure}[th!]
\includegraphics[width=1\linewidth]{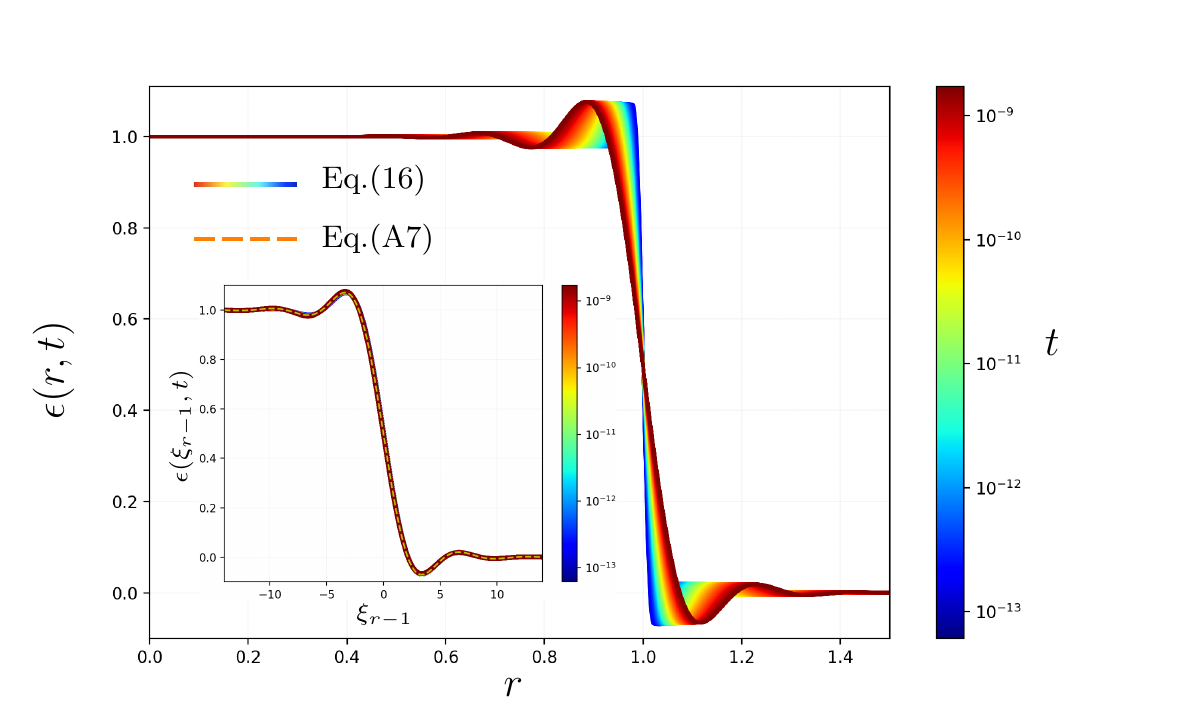}
\caption{Early-time film profiles for a stepped axisymmetric initial profile $\epsilon(r,0)=\mathcal{H}(1-r)$, as calculated form Eq.~\eqref{eq:convolution_axi} at several times $t$. The inset shows the same data as a function of the similarity variable $\xi_{r-1}=(r-1)t^{-1/6}$. The dashed line corresponds to the solution (see Eq.~\eqref{eq:1D_step}) of the linearized one-dimensional bending-driven levelling thin-film equation, with a stepped initial profile.}
\label{fig:early_time}
\end{figure}

From the profiles in Fig.~\ref{fig:exact_solution}b, we see that the early-time dynamics depends on the smoothness of
the initial condition. Indeed, in the left panel one observes an early crossover towards the attractor for a smooth initial condition, while in the right panel the crossover occurs later for a stepped axisymmetric initial condition. To investigate the latter case further, we plot in Fig.~\ref{fig:early_time} the raw solution (see Eq.~\eqref{eq:convolution_axi}) at early times, using the initial condition $\epsilon_0(r)=\mathcal{H}(r-1)$.
In the main figure, we observe that the elastic deformations are located close to the step front, which broadens with time. The evolution
is reminiscent of the self-similar one-dimensional-like capillary leveling reported previously for steps~\cite{Salez2012a}, as well as at early times for trenches~\cite{Baumchen2013} and holes~\cite{Backholm2014}. We therefore assume here too the existence of an early-time one-dimensional-like
self-similar regime, and introduce the natural similarity variable $\xi_{r-1}= (r-1)t^{-1/6}$ for our problem.
As shown in the inset, this rescaling allows us to collapse the early-time profiles onto a single curve.

For further analysis, we derive in the appendix the solution of the linearized one-dimensional bending-driven levelling thin-film equation, with a stepped initial profile. It is provided in Eq.~\eqref{eq:1D_step} and plotted for comparison in the inset of Fig.~\ref{fig:early_time}. The agreement is perfect, thus confirming the existence of the early-time one-dimensional-like self-similar regime for stepped axisymmetric initial conditions. Naturally, like for trenches~\cite{Baumchen2013} and holes~\cite{Backholm2014} in the capillary case, as the stepped early-time solution broadens, the opposing fronts eventually meet resulting in a crossover towards the universal attractor, including a decay of the central height. As such, the crossover time is expected to depend on the width of the initial profile. In the following section, we study generally the convergence time towards the universal attractor for any summable initial condition.

\subsection*{Universal attractor and convergence time}
By invoking the self-similarity of the Green's function (see Eq.~\eqref{ss}), the solution (see Eq.~\eqref{eq:convolution_axi}) of Eq.~\eqref{eq:ehd_linear} for an axisymmetric initial profile becomes:
\begin{equation}
\epsilon(\xi,t) = \frac{\mathcal{H}(t)}{t^{1/3}} \int \textrm{d}r'\, r' \epsilon_0(r')\int \textrm{d}\theta\, f\left(\sqrt{\xi^2+(r't^{-1/6})^2 - 2\xi r't^{-1/6}\cos(\theta)}\right)\ .
\end{equation}
At long positive times, this expression is equivalent to:
\begin{equation}
\epsilon(\xi,t) \simeq \frac{V_0}{t^{1/3}}\,f(\xi)\ ,
\label{eq:late_time_limit}
\end{equation}
where $V_0 = 2\pi\int \textrm{d}r\, r\, \epsilon _0(r)$ is the dimensionless volume of the initial perturbation profile. Therefore, the rescaled solution $\epsilon(\xi,t)/\epsilon(0,t) $ converges towards the self-similar attractor $f(\xi)/f(0)$, no matter the axisymmetric initial perturbation profile (provided it is summable), as previously suggested by Fig.~\ref{fig:exact_solution}b. For the sixth-order bending-driven thin-film equation, we thus find the intermediate asymptotic solution~\cite{barenblatt1996scaling} of the linearized problem to be the rescaled Green's function, which is reminiscent of the fourth-order capillary case~\cite{benzaquen2013intermediate}.

\begin{figure}[th!]
\begin{subfigure}[b]{0.52\linewidth}
\includegraphics[width=1\linewidth]{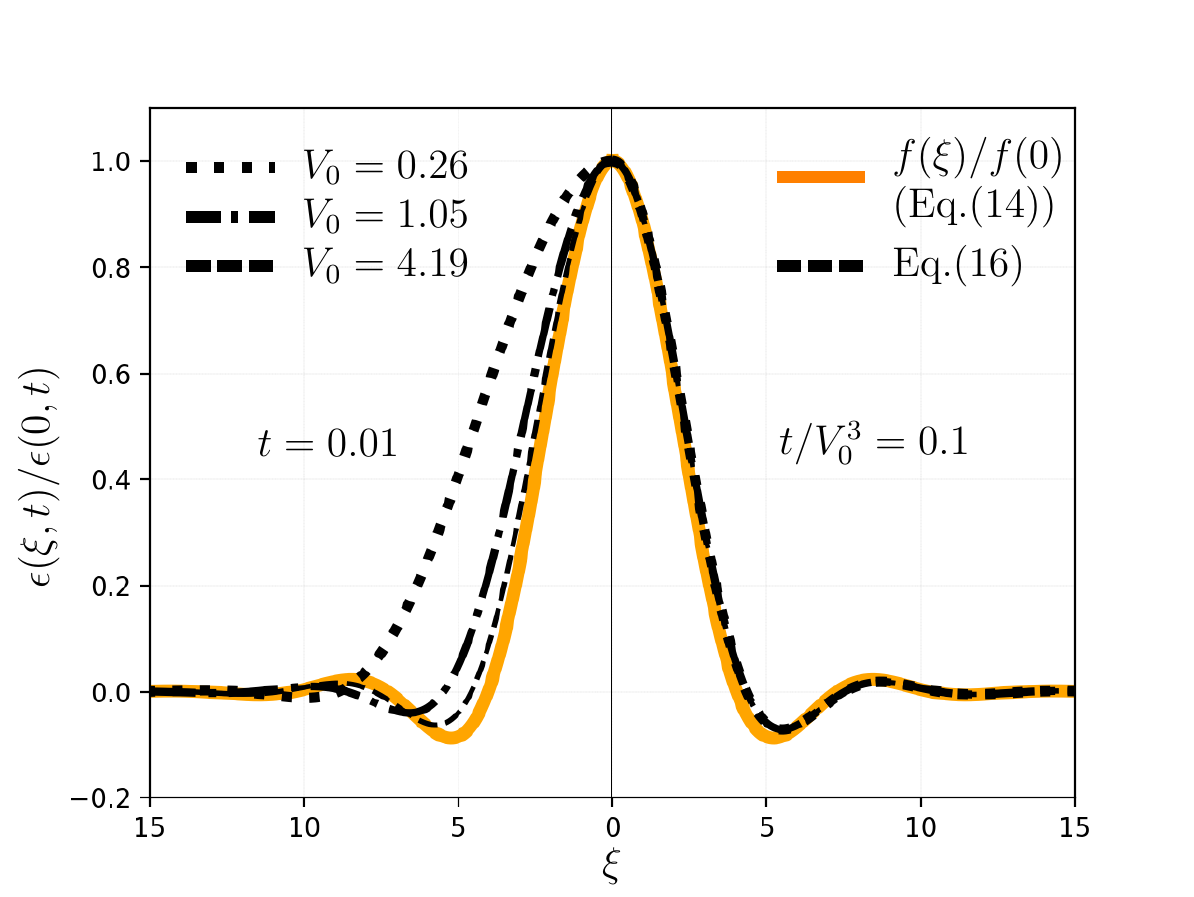}
\caption{}
\end{subfigure}
\begin{subfigure}[b]{0.52\linewidth}
\includegraphics[width=1\linewidth]{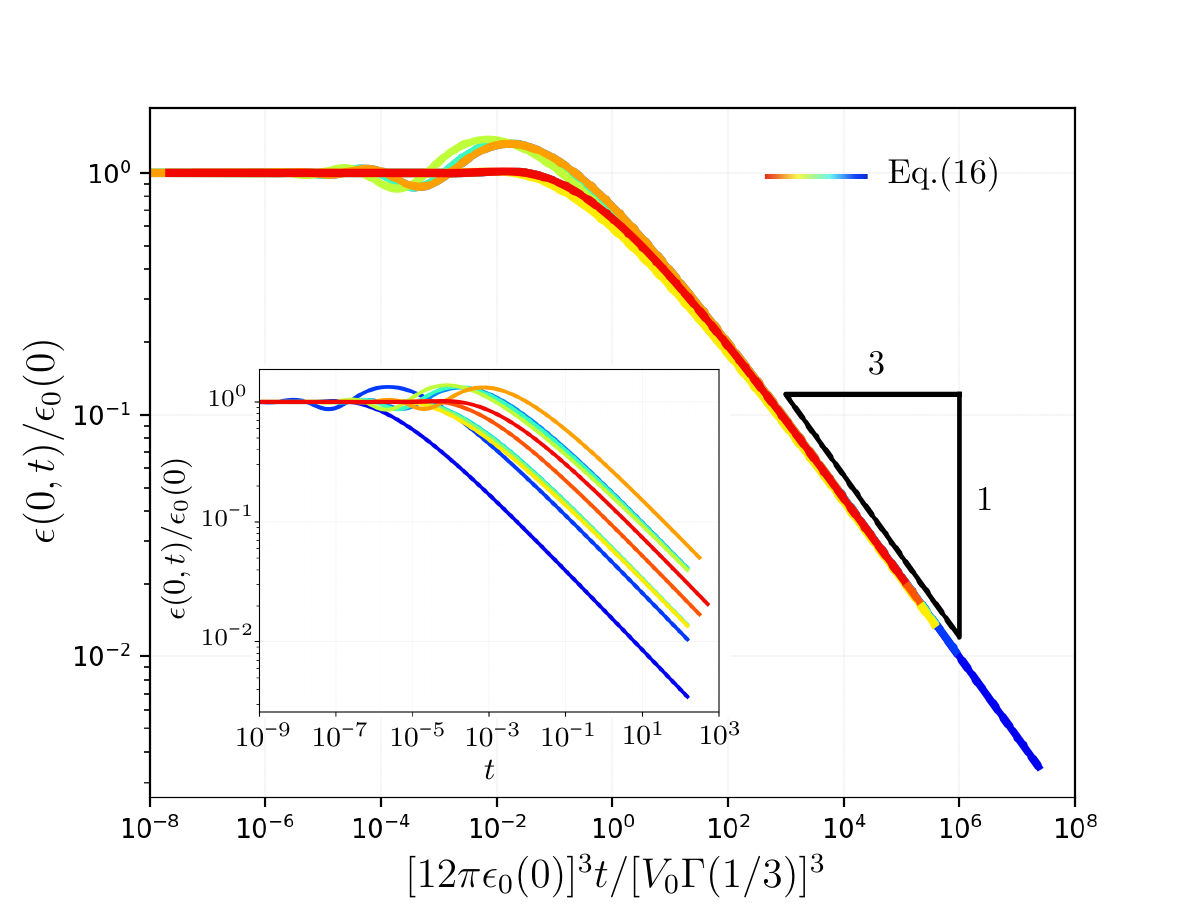}
\caption{}
\end{subfigure}
\caption{(a) (left) Rescaled solutions $\epsilon(\xi,0.01)/\epsilon(0,0.01)$ at a given time $t=0.01$ (dashed and dotted lines), for stepped axisymmetric initial profiles $\epsilon_0(r) = \left[1-(r/r_0)^2\right]^2\mathcal{H}(r_0-r)$, with $r_0 = 1, 2, 4$ giving the indicated values of $V_0$, as numerically evaluated from Eq.~\eqref{eq:convolution_axi}. (right) Rescaled solutions $\epsilon\left(\xi,0.1V_0^3\right)/\epsilon\left(0,0.1V_0^3\right)$, \textit{i.e.} at time $t=0.1V_0^3$ (dashed and dotted lines), for the three same initial profiles as in left panel, as numerically evaluated from Eq.~\eqref{eq:convolution_axi}. For comparison, on both the left and right panels, the self-similar attractor $f(\xi)/f(0)$ (see Eq.~\eqref{eq:attractor}) is shown. (b) Normalized central height of the perturbation (obtained from Eq.~\eqref{eq:convolution_axi}) as a function of the rescaled time $t/t_{\textrm{c}}$ (see Eq.~\eqref{eq:convergence_time}), for  initial profiles $\epsilon_0(r) =\left[1-(r/r_0)^2\right]^2\mathcal{H}(r_0-r)$ and $\epsilon_0(r) = \left[1-(r/r_0)^2\right]^2\mathcal{H}(r_0-r)$, with $r_0\in[0.5,0.75,1.0,1.5,2.0]$. The inset shows the same data as a function of time $t$, with $V_0/\epsilon_0(0)$ increasing from left to right.}
\label{fig:normalized_profile}
\end{figure}

As seen in Fig.~\ref{fig:exact_solution}b, and as discussed in the previous section on the early-time dynamics for stepped initial profiles, the time it takes for an arbitrary axisymmetric initial profile to converge to the self-similar attractor seems not to be unique. To investigate the role of the dimensionless volume $V_0$ of the perturbation on the convergence dynamics, we numerically evaluate from Eq.~\eqref{eq:convolution_axi} the solution $\epsilon(r,0.01)$ at a given time $t=0.01$, for axisymmetric initial profiles $\epsilon_0(r) = \left[1-(r/r_0)^2\right]^2\mathcal{H}(r_0-r)$, with $r_0 = 1, 2, 4$, giving $V_0 = 0.26, 1.05, 4.19$, respectively. The rescaled results are plotted in the left panel of Fig.~\ref{fig:normalized_profile}a, where we see that the three profiles do not collapse with one another.
This indicates that the dynamics is influenced by $V_0$. In contrast, when we numerically evaluate from Eq.~\eqref{eq:convolution_axi} the solutions $\epsilon\left(r,0.1V_0^3\right)$ at times $t=0.1V_0^3$, for the three same initial profiles as above, the rescaled profiles now seem to collapse with one another. This suggests that the dimensionless convergence time $t_{\textrm{c}}$ is proportional to $V_0^3$.

To quantitatively define $t_{\textrm{c}}$, we need a relevant criterion. A natural approach is to define a normalized mathematical distance between the solution and the attractor, and to fix some arbitrary but small upper bound to it. However, for regular axisymmetric initial profiles, convergence is typically occurring when the central height $\epsilon_0(0)$ of the initial perturbation profile matches the central height $\epsilon(0,t)\simeq V_0f(0)/t^{1/3}$ of the asymptotic solution (see Eq.~\eqref{eq:late_time_limit}). By using this simple criterion, combined with Eq.~\eqref{eq:attractor}, allows us to define the convergence time as:
\begin{equation}
t_{\textrm{c}} = \left[\frac{V_0\Gamma(1/3)}{12\pi\epsilon_0(0)} \right]^3\ .
\label{eq:convergence_time}
\end{equation}
Apart from numerical prefactors, the dimensionless convergence time is thus proportional to $[V_0/\epsilon_0(0)]^3$ only, confirming in particular the observation made in the right panel of Fig.~\ref{fig:normalized_profile}a for the $\epsilon_0(0)=1$ case.

In order to verify Eq.~\eqref{eq:convergence_time}, we compute from Eq.~\eqref{eq:convolution_axi} the central height $\epsilon(0,t)$ of the perturbation as a function of time, for various initial profiles. Specifically, we employ the homogeneous polynomials and axisymmetric step functions previously introduced, with $\epsilon_0(0)=1$ and $r_0\in[0.5,0.75,1.0,1.5,2.0]$. The results are plotted in Fig.~\ref{fig:normalized_profile}b, where the inset shows the normalized central height of the perturbation as a function of time, with $V_0/\epsilon_0(0)$ increasing from left to right. In the main figure, we show the same data but as a function of the rescaled time $t/t_{\textrm{c}}$, according to Eq.~\eqref{eq:convergence_time}. Apart from minor transient differences, related to the detailed shapes of the initial profiles (see previous section), the data essentially collapses onto a single curve. This collapse demonstrates that the convergence time is typically controlled by the initial area $V_0/\epsilon_0(0)$, but not by finer shape details of the initial profile. In addition, for rescaled times larger than unity, we recover the $1/3$ power law predicted by Eq.~\eqref{ss}.

In order to get practical insights, we close this section by returning to dimensional variables. By introducing the dimensional typical lateral width $\lambda$ of the initial perturbation, one can estimate the dimensionless volume as $V_0\sim \epsilon_0(0)(\lambda/h_0)^2$. Hence, the dimensional convergence time is given by $\sim\mu \lambda^6/(B h_0^3)$, as expected from the scaling analysis of Eq.~\eqref{eq:ehd_dimensional}. As such, the convergence is slower for a larger liquid film viscosity $\mu$ and a larger width $\lambda$ of the perturbation, while it is faster for a larger bending rigidity $B$ of the elastic plate and a larger liquid film thickness $h_0$.

\subsection*{Hydrodynamic fields and elastic energy}

With the attractor solution at hand (see Eqs.~\eqref{eq:attractor}~and~\eqref{eq:late_time_limit}), we can now derive the hydrodynamic fields in the long-term universal regime. Since these fields are of interest for experiments, we provide them using dimensional variables. Expressing Eq.~\eqref{eq:pressure} in cylindrical coordinates thanks to the axisymmetric film profile, making the derivatives with respect to the similarity variable $\xi$, and invoking Eq.~\eqref{eq:late_time_limit}, we find the pressure:

\begin{equation}
\begin{split}
\bar{p} &= 12^{2/3}\left(\frac{\mu^2 B}{h_0^3 \bar{t}^{\,2}}\right)^{1/3}\xi^{-1}\left\{\xi\left[\xi^{-1}\left(\xi \epsilon'\right)' \right]'\right\}'\\
  &\simeq \frac{12\mu \bar{V}_0}{h_0^3 \bar{t}}\xi^{-1}\left\{\xi\left[\xi^{-1}\left(\xi f'\right)' \right]'\right\}'\ ,
\end{split}
\label{eq:pressure_xi}
\end{equation}
with $\bar{V}_0=h_0^3V_0$ and where the prime $'$ indicates the derivative with respect to $\xi$. The first line in Eq.~\eqref{eq:pressure_xi} is valid at all times and expressed in terms of a specific solution $\epsilon$. As expected, the pressure decreases with time and liquid film thickness, and increases with the viscosity and bending rigidity. The second line in Eq.~\eqref{eq:pressure_xi} is valid at long times and expressed in terms of the universal attractor $f$. Therein, the different and perhaps non-intuitive prefactor appears due to the constitutive rescaling (see Eq.~\eqref{eq:late_time_limit}) inducing a lack of temporal decay in the attractor's amplitude (see Fig.~\ref{fig:normalized_profile}a). However, invoking the expression of $f$ in Eq.~\eqref{eq:attractor}, this second line of Eq.~\eqref{eq:pressure_xi} provides the long-term pressure field without the exact knowledge of any specific solution.

We now insert the obtained pressure of Eq.~\eqref{eq:pressure_xi} into Eq.~\eqref{eq:velocity} in order to obtain the velocity component along $\bar{r}=h_0 r$:

\begin{equation}
\begin{split}
\bar{u}_{\bar{r}} &= \frac{12^{5/6}}{2}\left(\frac{B h_0^3}{\mu \bar{t}^{\,5}}\right)^{1/6}\left[\frac{\bar{z}^{\,2}}{h_0^2}-\frac{\bar{z}}{h_0}\right]\left(\xi^{-1}\left\{\xi\left[\xi^{-1}\left(\xi \epsilon'\right)' \right]'\right\}'\right)'\\
    &\simeq \frac{12^{7/6}}{2}\left(\frac{\mu \bar{V}_0^{\,6}}{B h_0^{9}\bar{t}^{\,7}}\right)^{1/6}\left[\frac{\bar{z}^{\,2}}{h_0^2}-\frac{\bar{z}}{h_0}\right]\left(\xi^{-1}\left\{\xi\left[\xi^{-1}\left(\xi f'\right)' \right]'\right\}'\right)'\ .
\end{split}
\label{eq:velocity_xi}
\end{equation}
The first line in Eq.~\eqref{eq:velocity_xi} is valid at all times and expressed in terms of a specific solution $\epsilon$. As expected, the velocity decreases with time and viscosity, but increases with bending rigidity and liquid film thickness. The second line in Eq.~\eqref{eq:velocity_xi} is valid at long times and expressed in terms of the universal attractor $f$. Invoking the expression of $f$ in Eq.~\eqref{eq:attractor}, it provides the long-term radial velocity field without the exact knowledge of any specific solution.

By integrating the radial velocity of Eq.~\eqref{eq:velocity_xi} over the vertical coordinate $\bar{z}$ from $0$ to $h_0$ (at leading order), and by multiplying by the perimeter $2\pi\bar{r}$, we can express the radial volume flux in the film:

\begin{equation}
\begin{split}
\bar{Q}&= -\frac{2\pi}{12^{1/3}}\left(\frac{Bh_0^6}{\mu \bar{t}^{\,2}}\right)^{1/3}\xi\left(\xi^{-1}\left\{\xi\left[\xi^{-1}\left(\xi \epsilon'\right)' \right]'\right\}'\right)'\\
        &\simeq -2\pi\frac{\bar{V}_0}{\bar{t}}\xi\left(\xi^{-1}\left\{\xi\left[\xi^{-1}\left(\xi f'\right)' \right]'\right\}'\right)'\ .
\end{split}
\label{eq:flux_xi}
\end{equation}
The first line in Eq.~\eqref{eq:flux_xi} is valid at all times and expressed in terms of a specific solution $\epsilon$. As expected, the flux decreases with time and viscosity, but increases with bending rigidity and liquid film thickness. The second line in Eq.~\eqref{eq:flux_xi} is valid at long times and expressed in terms of the universal attractor $f$. Invoking the expression of $f$ in Eq.~\eqref{eq:attractor}, it provides the long-term radial volume flux without the exact knowledge of any specific solution.

Finally, after having obtained the local hydrodynamic fields, we consider a global quantity: the elastic energy of the plate. At small slopes, it is given by~\cite{landau1959course}:

\begin{equation}
\bar{F} = B\pi\int  \textrm{d}\bar{r}\, \bar{r}\left[\bar{r}^{-1}\partial_{\bar{r}}(\bar{r}\partial_{\bar{r}} \bar{h})\right]^2\ .
\end{equation}
Invoking the similarity variable $\xi$ as well as Eq.~\eqref{eq:late_time_limit}, the latter equation becomes:

\begin{equation}
\begin{split}
F &= 12^{1/3}\pi\left(\frac{\mu B^2 h_0^3}{\bar{t}}\right)^{1/3} \int \textrm{d}\xi\,\xi ^{-1}\left(\xi \epsilon'\right)'^2\\
  &\simeq \frac{\pi\mu \bar{V}_0^{\,2}}{h_0^3 \bar{t}}\ ,
\end{split}
\label{eq:energy}
\end{equation}
where we numerically estimated the integral $\int \textrm{d}\xi\,\xi ^{-1}\left(\xi f'\right)'^2\approx1/12$. The first line in Eq.~\eqref{eq:energy} is valid at all times and expressed in terms of a specific solution $\epsilon$. The second line in Eq.~\eqref{eq:energy} provides the long-term elastic energy without the exact knowledge of any specific solution. Importantly, for any initial condition, the long-term elastic energy evolves as the inverse of time, with a prefactor that only contains the viscosity, the volume of the perturbation and the liquid film thickness.

\section*{Extension to non-linear dynamics}
Being able to evaluate the typical time for convergence towards the self-similar attractor of the bending-driven thin-film equation can be crucial when quantitatively describing natural, biological or engineering processes, as well as model experimental systems~\cite{pedersen2019asymptotic}. However, so far, we have limited the analysis to the linearized problem. In the following, we re-examine the convergence to the self-similar attractor in the non-linear case described by Eq.~\eqref{eq:ehd_nondimensional}.

\begin{figure}[th!]
\centering
\includegraphics[width=1\linewidth]{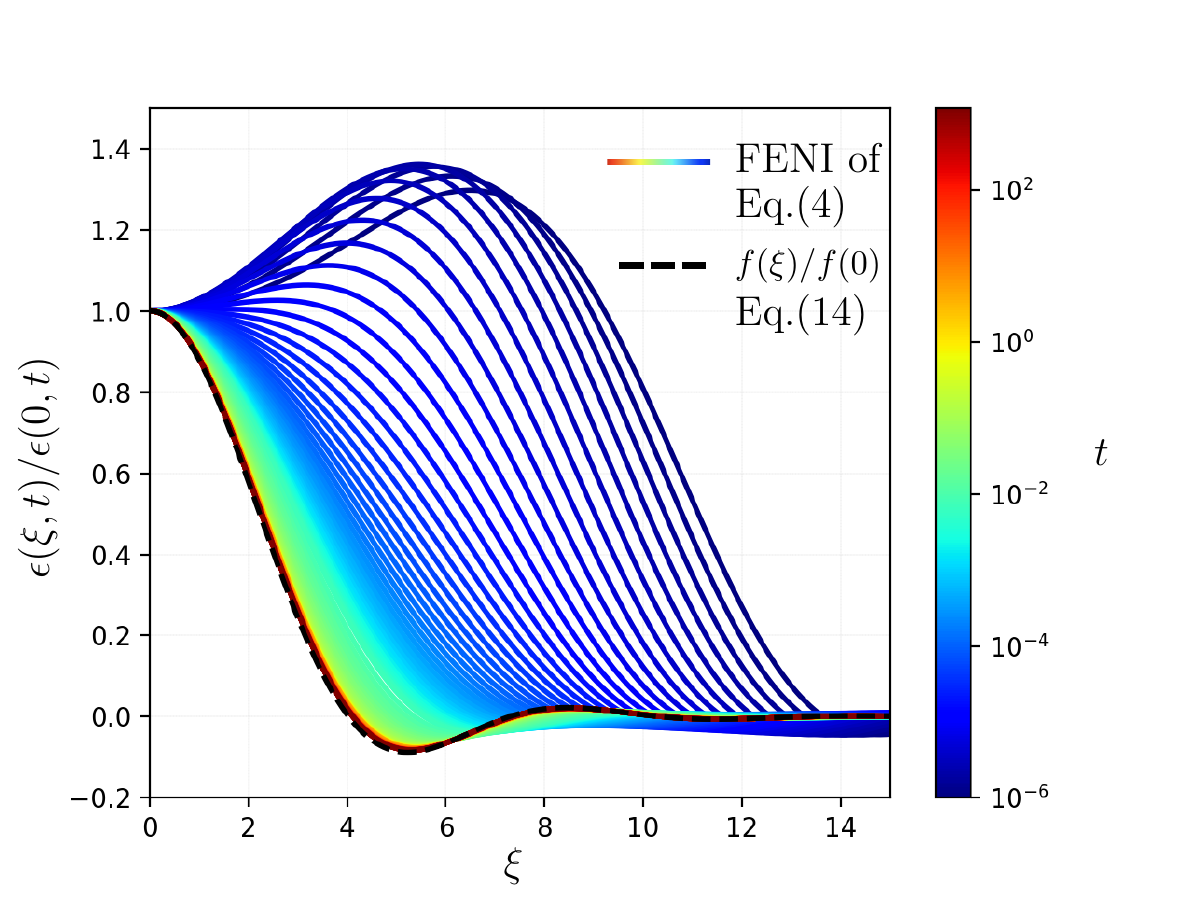}
\caption{Rescaled solution $\epsilon(\xi,t)/\epsilon(0,t)$ as a function of similarity variable $\xi=rt^{-1/6}$ (solid lines), for a stepped axisymmetric initial profile $\epsilon_0(r) = \mathcal{H}(1-r)$, at various times $t$ (color bar), as obtained from finite-element numerical integration (FENI) of Eq.~\eqref{eq:ehd_nondimensional}~\cite{pedersen2019asymptotic}. For comparison, the self-similar attractor $f(\xi)/f(0)$ (see Eq.~\eqref{eq:attractor}) is shown (dashed line).}
\label{fig:non-linear}
\end{figure}

We solve Eq.~\eqref{eq:ehd_nondimensional}, using a finite-element numerical integration (FENI)~\cite{pedersen2019asymptotic}. The rescaled solution $\epsilon(\xi,t)/\epsilon(0,t)=[h(\xi,t)-1]/[h(0,t)-1]$, for a stepped initial axisymmetric profile $\epsilon_0(r) =h(r,0)-1= \mathcal{H}(1-r)$ is shown in Fig.~\ref{fig:non-linear} for various times $t$.

We observe that the rescaled non-linear solution converges in time towards the self-similar attractor $f(\xi)/f(0)$ (see Eq.~\eqref{eq:attractor}) of the linear case. We have checked (not shown) that this statement holds for all the various compact-support axisymmetric initial profiles tested, and are thus led to conjecture its validity for any summable initial profile of arbitrary magnitude. The physical reason behind this phenomenon is rooted in the dissipative character of Eq.~\eqref{eq:ehd_nondimensional}, that ensures the condition $\epsilon(r,t) \ll 1$ to be always reached eventually, at sufficiently long times.

\section*{Conclusion}
We studied the sixth-order bending-driven thin-film equation, both theoretically and numerically. We derived the Green's function of the linearized problem, and showed that it represents a universal self-similar attractor. As such, the linear solution from any summable axisymmetric initial perturbation profile converges towards the rescaled Green's function at intermediate times. In addition, for stepped axisymmetric initial conditions, we demonstrated the existence of an early-time one-dimensional-like self-similar regime. Besides, we characterized the convergence time towards the long-term universal attractor in terms of the relevant physical and geometrical parameters, and we derived the hydrodynamic fields and elastic energy in the universal regime. Finally, we extended numerically our analysis to the non-linear case, and verified the convergence towards the self-similar attractor.

\section*{Acknowledgments}
The authors thank Tak Shing Chan and Jacco Snoeijer for valuable discussions. The authors acknowledge funding from the Research Council of Norway (grant number $263056$), and from the Agence Nationale de la Recherche (ANR-21-ERCC-0010-01 \textit{EMetBrown}). 

\section*{Appendix: 1D linearized problem}
\renewcommand{\theequation}{A\arabic{equation}}
\setcounter{equation}{0}
We derive the Green's function, the associated self-similar attractor, and the convergence time for the 1D version of Eq.~\eqref{eq:ehd_linear}. The linear differential operator is now $\mathcal{L}_{\textrm{1D}} = \partial_t - \partial_x^6$, with $x$ the single spatial coordinate. We start by taking the Fourier transform of Eq.~\eqref{eq:Greens} in 1D, which yields:
\begin{equation}
\hat{G}_{\textrm{1D}}(k,\omega) = \frac{1}{i\omega + k^6}\ .
\end{equation}
From the inverse Fourier transform and the residue theorem, we get:
\begin{equation}
G_{\textrm{1D}}(x,t) = \frac{\mathcal{H}(t)}{2\pi}\int \textrm{d}k\, e^{-k^6t}e^{ikx}\ .
\end{equation}
Using the similarity variable $\zeta = xt^{-1/6}$, it follows:
\begin{equation}
G_{\textrm{1D}}(\zeta,t) = \frac{\mathcal{H}(t)}{t^{1/6}}f(\zeta)\ ,
\label{eq:greens1D}
\end{equation}
with:
\begin{equation}
\begin{split}
f_{\textrm{1D}}(\zeta) = &-\frac{2}{\Gamma\left(-\frac{1}{6}\right)} {}\,_0F_4\left( \left\{\frac{1}{3}, \frac{1}{2}, \frac{2}{3}, \frac{5}{6} \right\}, -\left( \frac{\zeta}{6} \right)^6  \right)\\ &- \frac{\zeta^2}{12\sqrt{\pi}}{}\,_0F_4\left( \left\{ \frac{2}{3}, \frac{5}{6}, \frac{7}{6}, \frac{4}{3} \right\}, -\left( \frac{\zeta}{6}  \right)^6  \right)\\ &+ \frac{\zeta^4}{432\,\Gamma\left(\frac{7}{6}\right)} {}\,_0F_4\left( \left\{ \frac{7}{6}, \frac{4}{3}, \frac{3}{2}, \frac{5}{3} \right\}, -\left( \frac{\zeta}{6} \right)^6  \right)\ ,
\end{split}
\label{eq:attractor1D}
\end{equation}
where $\Gamma$ is the gamma function and $_0F_4$ is the (0,4)-hypergeometric function.

The solution is then obtained from the 1D convolution with the initial profile:
\begin{equation}
\epsilon(x,t) = \int \textrm{d}x'\, G_{\textrm{1D}}(x-x',t)\epsilon_0(x')\ .
\label{eq:convolution1D}
\end{equation}
Using the similarity variable $\zeta = xt^{-1/6}$ and Eq.~\eqref{eq:greens1D}, Eq.~\eqref{eq:convolution1D} becomes:
\begin{equation}
\epsilon(x,t) = \frac{\mathcal{H}(t)}{t^{1/6}} \int \textrm{d}x'\, \epsilon_0(x')f_{\textrm{1D}}(\zeta-x't^{-1/6})\ .
\label{convss}
\end{equation}
For the particular case of a stepped initial condition $\epsilon_0(x)=\mathcal{H}(-x)$, Eq.~\eqref{convss} reduces to:

\begin{equation}
\begin{split}
\epsilon(\zeta) = \frac{1}{2} + \frac{\zeta}{2\pi}&\left[-2\Gamma\left(\frac{7}{6}\right) {}\,_1F_5\left( \left\{\frac{1}{6}\right\}, \left\{\frac{1}{3}, \frac{1}{2}, \frac{2}{3}, \frac{5}{6}, \frac{7}{6} \right\}, -\left( \frac{\zeta}{6} \right)^6\right)\right.  \\ &+ \left.\frac{\sqrt{\pi}\zeta^2}{18}{}\,_1F_5\left(\left\{\frac{1}{2}\right\}, \left\{ \frac{2}{3}, \frac{5}{6}, \frac{7}{6}, \frac{4}{3}, \frac{3}{2} \right\}, -\left( \frac{\zeta}{6}  \right)^6  \right)\right.\\ &+ \left.\frac{\zeta^4\Gamma\left(-\frac{1}{6}\right)}{2160} {}\,_1F_5\left( \left\{\frac5{6}\right\}, \left\{ \frac{7}{6}, \frac{4}{3}, \frac{3}{2}, \frac{5}{3}, \frac{11}{6} \right\}, -\left( \frac{\zeta}{6} \right)^6  \right) \right]\ .
\end{split}
\label{eq:1D_step}
\end{equation}
Besides, at long positive times, Eq.~\eqref{convss} is equivalent to:
\begin{equation}
\epsilon(\zeta,t) \simeq \frac{A_0}{t^{1/6}}\,f_{\textrm{1D}}(\zeta)\ ,
\label{eq:late_time_limit1D}
\end{equation}
where $A_0=\int \textrm{d}x'\, \epsilon_0(x')$ is the dimensionless area of the initial perturbation profile. Therefore, the rescaled solution $\epsilon(\zeta,t)/\epsilon(0,t) $ converges towards the universal self-similar attractor $f_{\textrm{1D}}(\zeta)/f_{\textrm{1D}}(0)$, no matter the initial perturbation profile (provided it is summable).

Assuming as a criterion that the convergence to the universal attractor is typically occurring when the central height $\epsilon_0(0)$ of the initial perturbation profile matches the central height $\epsilon(0,t)\simeq A_0\,f_{\textrm{1D}}(0)/t^{1/6}$ of the asymptotic solution (see Eq.~\eqref{eq:late_time_limit1D}), we find the 1D convergence time:
\begin{equation}
t_{\textrm{c,1D}} = \left[\frac{2A_0}{\epsilon_0(0)\Gamma(-1/6)} \right]^6 \ .
\end{equation}

\bibliographystyle{ieeetr}
\bibliography{Pedersen2020}
\end{document}